\documentclass[aps,prc,superscriptaddress,twoside,twocolumn,nofootinbib,%
showpacs,floatfix]{revtex4-1}

\usepackage{amsmath,amssymb}
\usepackage{graphicx,bm}
\usepackage{slashed}

\usepackage{ulem} %% for strike-through
\usepackage[usenames]{color}

\renewcommand\sout{\bgroup \color{red} \ULdepth=-.5ex \ULset}

\begin{document}

\title{\boldmath Role of high-spin hyperon resonances
in the reaction of $\gamma p \to K^+ K^+ \Xi^-$}

\author{J. Ka Shing Man}

\affiliation{Department of Physics and Astronomy, University of Georgia,
Athens, GA 30602, USA}

\author{Yongseok Oh}%
\email{yohphy@knu.ac.kr}

\affiliation{Department of Physics, 
Kyungpook National University, Daegu 702-701, Korea}
\affiliation{Excited Baryon Analysis Center, Thomas Jefferson National Accelerator Facility, 
Newport News, VA 23606, USA}
\affiliation{Asia Pacific Center for Theoretical Physics, 
POSTECH, Pohang 790-784, Korea}

\author{K. Nakayama}%
\email{nakayama@uga.edu}

\affiliation{Department of Physics and Astronomy, University of Georgia,
Athens, GA 30602, USA}
\affiliation{Institut f\"ur Kernphysik, Forschungszentrum
J\"ulich, D-52425 J\"ulich, Germany}

\date{}

\begin{abstract}
The recent data taken by the CLAS Collaboration at the Thomas Jefferson
National Accelerator Facility for the reaction of
$\gamma p \to K^+ K^+ \Xi^-$ are reanalyzed within a relativistic meson-exchange 
model of hadronic interactions. 
The present model is an extension of the one developed in an earlier work by 
Nakayama, Oh, and Haberzettl [Phys. Rev. C 74, 035205 (2006)]. 
In particular, the role of the spin-5/2 and -7/2 hyperon resonances,
which were not included in the previous model, is investigated in the
present study.
It is shown that the contribution of the $\Sigma(2030)$ hyperon having
spin-$7/2$ and positive parity has a key role to bring the model
predictions into a fair agreement with the measured data for the
$K^+\Xi^-$ invariant mass distribution.    
\end{abstract}

\pacs{25.20.Lj, % Photoproduction reactions
      13.60.Le, % Meson production
      13.60.Rj  % Baryon production
      14.20.Jn  % Hyperons
      } %

\maketitle

%%%%%%%%%%%%%%%%%%%%%%%%%%%%%%%%%%%%%%%%%%%%%%%%%%%%%%%%%%%%%
\section{Introduction}

While there are extensive studies on the production of $\Lambda$ and
$\Sigma$ hyperons~\cite{CLAS05c,LEPS03b,SAPHIR98,TTWF07}, only a few
attempts were made so far to understand the production of
multi-strangeness hyperons.
In addition, the spectrum of multi-strangeness hyperons is yet to be
explored.
For example, if the flavor SU(3) symmetry is good for the classification
of baryons, we expect to have the number of $\Xi$ resonances being equal
to the sum of the number of nucleon resonances and the number of $\Delta$
resonances. 
However, while one can find more than 20 nucleon resonances and more 
than 20 $\Delta$ resonances in the review of particles compiled by 
the Particle Data Group (PDG), only a dozen or so $\Xi$'s have been
identified until today~\cite{PDG10}.
Furthermore, only two of them, $\Xi(1318)$ and $\Xi(1530)$, have four-star
status according to the PDG.
One of the reasons for this situation is that $\Xi$ hyperons, being
particles with strangeness $S=-2$, are difficult to produce as they have
relatively small production rates. 
Namely, they can only be produced via indirect processes from the nucleon, 
which leads to very small production cross sections.
It is, then, natural to see that the case of $\Omega$ hyperons is even
worse.

The investigation on multi-strangeness baryons, however, is expected to
provide a very important tool to understand the structure of baryons and to
distinguish various phenomenological models on baryon mass spectrum.
In fact, most theoretical models on baryon spectrum could describe the
mass spectra of the ground states of baryon octet and decuplet.
However, this is basically a consequence of the group structure of the
flavor SU(3) symmetry and the symmetry breaking terms. 
On the other hand, these models give very different predictions on
the spectrum of excited states of hyperons.
This high model-dependence can be easily found in the spectrum of $\Xi$
resonances. (See, e.g., Ref.~\cite{Oh07}.)
In particular, description of the low-lying $\Xi$ baryons, $\Xi(1620)$ and
$\Xi(1690)$, would provide important insight to understanding baryon
structure.
Most quark models predict higher mass for the $\Xi(1690)$ and its
predicted spin-parity quantum numbers depend on the details of the
model~\cite{CIK81,CI86,PR07}.
Furthermore, the low mass of one-star-rated $\Xi(1620)$ is puzzling to
quark models.
On the other hand, these states were claimed to be dynamically
generated $S$-wave resonances in unitarized meson-baryon interaction
models~\cite{ROB02,GLN04}.
In the bound state approach to the Skyrme model, these
states are described as bound states of the soliton and two kaons, and,
therefore, they are analogous to the $\Lambda(1405)$~\cite{Oh07}.
Furthermore, in this model, the existence of the $\Xi(1620)$ is required
as it accompanies the $\Xi(1690)$.
Therefore, confirming the existence of $\Xi$ resonances and verifying
their quantum numbers will help shed light on the understanding of
multi-strangeness baryon structure.

Experimentally, there have been several recent reports on the new measurements
for the properties of $\Xi$ weak decays~\cite{KTeV05,NA48/1-07} and for the
$\Xi$ resonances~\cite{Belle02,STAR07}.
In particular, the BABAR Collaboration reported the evidence of the
spin-parity of the $\Xi(1690)$ being $\frac12^-$~\cite{BABAR08}.
The investigation to understand the production mechanism of the ground state
$\Xi$ baryon has been recently initiated by the CLAS Collaboration at the
Thomas Jefferson National Accelerator Facility
(TJNAF)~\cite{CLAS04d,PDGN05,CLAS06d}.
It has established the feasibility to investigate hyperon spectroscopy via
photoproduction reactions like $\gamma p \to K^+ K^+ \Xi^-$
and $\gamma p \to K^+ K^+ \pi^- \Xi^0$~\cite{PDGN05,CLAS04d}.
The first dedicated experiment for these reactions has been carried out
and the data for the total and differential cross
sections as well as the $K^+K^+$ and $K^+\Xi^-$ invariant mass
distributions for the $\gamma p \to K^+ K^+ \Xi^-$ reaction have been
reported in Ref.~\cite{CLAS07b}.
To our knowledge, this is the first data set measured for the
exclusive production of the $\Xi$ in photon-nucleon scattering.

Theoretical investigation of the reaction mechanism of $\Xi$ production
has been started only recently~\cite{LK03c,NOH06,OHN07-NOH07}.
The work of Ref.~\cite{LK03c} was devoted to search for the
pentaquark exotic $\Xi$ hyperon, and the detailed analysis for
the reaction of $\gamma N \to K K \Xi$ in the energy region up to
$\sim 4$~GeV based on the data obtained at the TJNAF~\cite{CLAS07b} was
initiated by the work of Ref.~\cite{NOH06}.
In the present work we extend the model of Ref.~\cite{NOH06} to address
the role of the contributions from the $\Lambda$ and $\Sigma$ hyperon
resonances of spin-$5/2$ and -$7/2$ in the production mechanism.

\begin{figure}[t!]\centering
\vskip -0.4cm
\includegraphics[width=0.4\textwidth,angle=0,clip]{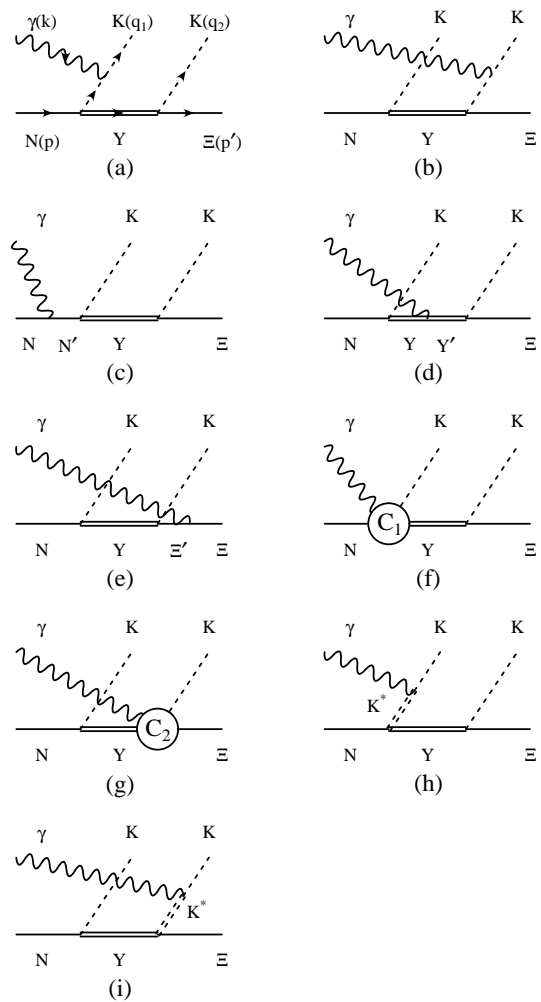}
\caption{\label{fig:diagram}
Diagrams contributing to the reaction mechanism of $\gamma N \to K K \Xi$.
The intermediate baryon states are denoted as $N'$ for the nucleon and
$\Delta$ resonances, $Y,Y'$ for the $\Lambda$ and $\Sigma$ resonances,
and $\Xi'$ for $\Xi(1318)$ and $\Xi(1530)$.
The intermediate mesons in the $t$-channel are $K$ [(a) and (b)] and
$K^*$ [(h) and (i)].
The diagrams (f) and (g) contain the generalized contact currents that
maintain gauge invariance of the total amplitude.
Diagrams corresponding to (a)--(i) with $K(q_1) \leftrightarrow K(q_2)$
are also understood.}
\end{figure}

The approach of Ref.~\cite{NOH06} is based on a relativistic meson-exchange
model of strong interactions and the reaction amplitude is calculated at
the tree-level approximation by making use of effective Lagrangian.
As was discussed in Ref.~\cite{NOH06}, the main contribution to the
reaction mechanisms of the $\gamma N \to K K \Xi$ reaction comes from the
excitations of intermediate hyperons because of the absence of exotic
mesons that have strangeness $S = +2$. (See  Fig.~\ref{fig:diagram}.)
This is the main difference from the reaction mechanism of the
production of kaon and antikaon pair, i.e., $\gamma N \to K \bar{K} N$,
which is dominated by the production of the $\phi$ meson~\cite{ONL04}.
Therefore, the reaction of $\Xi$ photoproduction would be a very useful
tool to investigate hyperons having $S = -1$.

%%%%%%%%%%%%%%%%%%%%%%%%%%%%%%%%%%%%%%%%%%%%%%
%    Figure 1
%
%%%%%%%%%%%%%%%%%%%%%%%%%%%%%%%%%%%%%%%%%%%%%%%%%

In fact, there are number of three- and four-star $\Lambda$ and $\Sigma$
hyperon resonances~\cite{PDG10} that may have crucial role in the mechanism
of $\Xi$ production in the photon-nucleon scattering.
Although their properties are not precisely determined yet, it has been
explicitly shown in Ref.~\cite{NOH06} that a potential contribution of hyperon
resonances with a mass around $2$~GeV can affect the $K^+\Xi^-$ invariant
mass distribution in a significant way.
The original model of Ref.~\cite{NOH06}, which includes only the lower mass
hyperon resonances of spin-$1/2$ and -$3/2$, predicted a double-bump
structure in the $K^+ \Xi^-$ invariant mass distribution that was not
exhibited by the observed data~\cite{CLAS06d}.
Instead, it was shown that a hyperon resonance of a mass around $2$~GeV can
fill the valley between the double-bump to explain the experimental data.
However, according to the PDG, there is no identified hyperon resonances of
spin-$1/2$ or -$3/2$ in this mass region, but such resonances 
are found to have spin-$5/2$ or -$7/2$~\cite{PDG10}.
(See Table.~\ref{tbl:hyperons}.)

%%%%%%%%%%%%%%%%%%%%%%%%%%%%%%%%%%%%%%%%%%%%%%
%    Table (I)
\begin{table*}[t]
\centering
\caption{The $\Lambda$ and $\Sigma$ hyperons listed by the Particle Data
Group~\cite{PDG10} as three-star or four-star states.
The decay widths and branching ratios of high-mass resonances
$m_Y^{} > 1.6$ GeV are in a broad range, and
the coupling constants are determined from their central values.}
\begin{tabular}{c@{\extracolsep{1em}}cccc|ccccc} \hline\hline
\multicolumn{5}{c|}{$\Lambda$ states} & \multicolumn{5}{c}{$\Sigma$ states} \\
\hline State & $J^P$ & $\Gamma$ (MeV) & Rating & $|g_{N\Lambda K}^{}|$
& State & $J^P$ & $\Gamma$ (MeV) & Rating  & $|g_{N\Sigma K}^{}|$ \\ \hline
$\Lambda(1116)$ & $1/2^+$ &  & **** &     &
  $\Sigma(1193)$ & $1/2^+$ &              & **** &     \\
$\Lambda(1405)$ & $1/2^-$ & $\approx 50$ & **** &       &
  $\Sigma(1385)$ & $3/2^+$ & $\approx 37$ & **** &      \\
$\Lambda(1520)$ & $3/2^-$ & $\approx 16$ & **** &       &
                 &         &              &     &       \\
\hline
$\Lambda(1600)$ & $1/2^+$ & $\approx 150$ & *** &  4.2 &
  $\Sigma(1660)$ & $1/2^+$ & $\approx 100$ & *** & 2.5  \\
$\Lambda(1670)$ & $1/2^-$ & $\approx 35$ & **** &  0.3 &
  $\Sigma(1670)$ & $3/2^-$ & $\approx 60$ & **** & 2.8 \\
$\Lambda(1690)$ & $3/2^-$ & $\approx 60$ & **** &  4.0 &
  $\Sigma(1750)$ & $1/2^-$ & $\approx 90$ & *** & 0.5 \\
$\Lambda(1800)$ & $1/2^-$ & $\approx 300$ & *** &  1.0 &
  $\Sigma(1775)$ & $5/2^-$ & $\approx 120$ & **** & \\
$\Lambda(1810)$ & $1/2^+$ & $\approx 150$ & *** &  2.8 &
  $\Sigma(1915)$ & $5/2^+$ & $\approx 120$ & **** & \\
$\Lambda(1820)$ & $5/2^+$ & $\approx 80$ & **** &      &
  $\Sigma(1940)$ & $3/2^-$ & $\approx 220$ & *** & $< 2.8$  \\
$\Lambda(1830)$ & $5/2^-$ & $\approx 95$ & **** &      &
  $\Sigma(2030)$ & $7/2^+$ & $\approx 180$ & **** & \\
$\Lambda(1890)$ & $3/2^+$ & $\approx 100$ & **** & 0.8 &
  $\Sigma(2250)$ & $?^?$ & $\approx 100$ & ***  & \\
$\Lambda(2100)$ & $7/2^-$ & $\approx 200$ & **** & & & & & \\
$\Lambda(2110)$ & $5/2^+$ & $\approx 200$ & *** & & & & & \\
$\Lambda(2350)$ & $9/2^+$ & $\approx 150$ & *** & & & & & \\
\hline\hline
\end{tabular}
\label{tbl:hyperons}
\end{table*}
%%%%%%%%%%%%%%%%%%%%%%%%%%%%%%%%%%%%%%%%%%%%%%

It is, therefore, necessary to explore the role of the four-star
spin-$5/2$ and -$7/2$ hyperons at a mass around $2$~GeV in $\Xi$
photoproduction by extending the model of Ref.~\cite{NOH06}.
It is the purpose of the present work, and here we concentrate on
the photoproduction mechanism of the ground-state $\Xi(1318)$ off the
proton target, i.e.,
$\gamma p \to K^+ K^+ \Xi^-$, where the data (including the $K^+\Xi^-$
invariant mass distribution) were collected by the CLAS
Collaboration~\cite{CLAS07b}.

In the next Section, we describe the extension of the original model of
Ref.~\cite{NOH06} for $\Xi$ photoproduction, whose dynamical content is
shown in Fig.~\ref{fig:diagram}.
We here include the contributions from the hyperon resonances of
spin-$5/2$ and -$7/2$.
Our results are then compared to the experimental data of the CLAS
Collaboration~\cite{CLAS07b} in Sec.~III, focusing, in particular, on the 
role of the higher spin intermediate hyperons.
Section~IV contains a summary and discussions.
Some details of the present model Lagrangian are given in the Appendix.

\section{Model for $\bm{\gamma p \to K^+K^+\Xi^-}$}

The model for describing the $\gamma p \to K^+K^+\Xi^-$ reaction in the
present work is essentially same as that of Ref.~\cite{NOH06}, which
consists of the amplitudes shown in Fig.~\ref{fig:diagram}.
The only difference is that, here we include the spin-$5/2$ and -$7/2$
hyperons as will be specified below.
Such high-spin resonances were not considered in the model of
Ref.~\cite{NOH06}.
In the following, for completeness, we first discuss the most relevant
features of the model of Ref.~\cite{NOH06}.

The three- and four-star hyperons which can contribute to the reaction of
$\Xi$ photoproduction are summarized in Table~\ref{tbl:hyperons}.
Among them, only for the low-mass resonances, i.e.,
$\Lambda(1116)$, $\Lambda(1405)$, $\Lambda(1520)$, $\Sigma(1190)$, and
$\Sigma(1385)$, we have sufficient information to determine the relevant
hadronic and electromagnetic coupling constants either from the experimental
data~\cite{PDG10} or from quark models with SU(3) symmetry consideration.
These hyperon resonance parameters and the estimation of the corresponding
coupling constants can be found in Ref.~\cite{NOH06}.

For higher-mass resonances (those listed in Table~\ref{tbl:hyperons} with 
a mass larger than $1.6$~GeV), however, there is no sufficient 
information to extract the necessary parameters,
in particular, their coupling constants to the $\Xi$.
The only available information relevant to the present reaction involving
the higher hyperon resonances are their partial decay widths into the
$N \bar{K}$ channel~\cite{PDG10}.
This information allows us to extract the magnitude of the corresponding
$NYK$ coupling constants, and their estimated values are displayed in
Table~\ref{tbl:hyperons}.
Therefore, in Ref.~\cite{NOH06}, only the diagrams (a)--(g) in
Fig.~\ref{fig:diagram} with $Y=Y'$ in (d), where the only additional parameter
is the $\Xi YK$ coupling constant, were considered, in addition to
restricting $Y$ to spin-$1/2$ and -$3/2$ hyperons.

It was, then, noted that, unless the $\Xi YK$ coupling constants are much
larger than the corresponding $NYK$ coupling constants, hyperon resonances with
$J^P=1/2^+$ and $3/2^-$ yield much smaller contributions to the reaction
amplitude when compared to the $J^P=1/2^-$ and $3/2^+$ resonance contributions.
As was pointed out in Ref.~\cite{NOH06}, this can be understood by
considering the intermediate hyperon resonance on its mass shell.
By analyzing the production amplitude with an intermediate hyperon
resonance on its mass shell, the photoproduction amplitude was found to
be proportional to either the sum or difference of the baryon masses and
energies depending on the spin-parity of the resonance, namely,
\footnote{The error committed in Eq.~(2) of Ref.~\cite{NOH06}
is corrected here. However, the results and the conclusions
of Ref.~\cite{NOH06} remain unchanged.} 
\begin{equation}
\begin{aligned}
|M_{1/2^\pm}|^2 & \propto  (\varepsilon_N^{} \mp m_N^{})
(\varepsilon_\Xi^{} \mp m_\Xi^{}),  \\
|M_{3/2^\pm}|^2 & \propto  (\varepsilon_N^{} \pm m_N^{})
(\varepsilon_\Xi^{} \pm m_\Xi^{}),
\end{aligned}\label{MMDF}
\end{equation}
where $M_{J^P}^{}$ denotes the photoproduction amplitude involving the
intermediate hyperon with the spin-parity $J^P$;
$\varepsilon_i\equiv \sqrt{\bm{p}_i^2 + m_i^2}$ with $\bm{p}_i^{}$ and 
$m_i^{}$ denoting the momentum and mass, respectively, of the nucleon
or the $\Xi$ as $i=N$ or $\Xi$. 
This proportionality is valid only when the intermediate hyperon lies on its
mass shell, and it does not quite apply to the low-mass resonances
which are far off-shell in the present reaction.
Among the $J^P=1/2^-$ and $3/2^+$ resonances, assuming the $\Xi YK$ coupling
strength to be of the order of the $NYK$ coupling strength, only the
$\Lambda(1800)1/2^-$ and $\Lambda(1890)3/2^+$ resonances contribute
significantly.%
\footnote{
For $\Lambda$ resonances, one can expect
$|g_{\Xi\Lambda K}^{}/g_{N\Lambda K}^{}| \le 1$ since
$g_{\Xi\Lambda K}^{} = g_{N\Lambda K}^{}$ for a singlet $\Lambda$ and
$g_{\Xi\Lambda K}^{}/g_{N\Lambda K}^{} = (1-4f)/(1+2f)$ for an octet
$\Lambda$~\cite{deS63}, which is less than 1 for $0<f<1$.}
Therefore, only these two high-mass resonances were considered in the
model of Ref.~\cite{NOH06}.%
\footnote{Among the $\Sigma$ resonances, the only candidate is
the $\Sigma(1750)$ with $J^P = 1/2^-$. However, it has a very small
coupling constant $g_{N\Sigma K}^{}$.}

In the present work, we include the contributions from the spin-$5/2$ and
-$7/2$ hyperon resonances to the model of Ref.~\cite{NOH06}, i.e.,
on top of the $\Lambda(1800)1/2^-$ and $\Lambda(1890)3/2^+$ resonances.
This is motivated by the observation that the measured $K^+\Xi^-$ invariant
mass distribution~\cite{CLAS07b} exhibits no double-humped structure
predicted by the model of Ref.~\cite{NOH06} at around $2$~GeV.
In fact, one can find four-star spin-$5/2$ and -$7/2$ hyperons of
this mass range in Table~\ref{tbl:hyperons} and these resonances should
be taken into account for understanding the production mechanism.
By performing the similar analysis which lead to Eq.~(\ref{MMDF}), we can
obtain the analogous features for spin-$5/2$ and -$7/2$ resonances in the
on-shell limit, which read
\begin{equation}
\begin{aligned}
|M_{5/2^\pm}|^2 & \propto  (\varepsilon_N^{} \mp m_N^{})
(\varepsilon_\Xi^{} \mp m_\Xi^{})  ,  \\
|M_{7/2^\pm}|^2 & \propto  (\varepsilon_N^{} \pm m_N^{})
(\varepsilon_\Xi^{} \pm m_\Xi^{})  .
\end{aligned}\label{MMDF57}
\end{equation}
This allows us to consider only the $J^P=5/2^-$ and $7/2^+$ resonances.
We then see from Table~\ref{tbl:hyperons} that the $\Sigma(2030)7/2^+$,
$\Sigma(1775)5/2^-$, and $\Lambda(1830)5/2^-$ are the only candidates.
Among them the $\Sigma(1775)5/2^-$, and $\Lambda(1830)5/2^-$ have a 
mass below or very close to the threshold. Since we are interested in the 
resonances of mass about 2~GeV, we focus on the role of the 
$\Sigma(2030)7/2^+$ resonances in the present work.
In fact, we have verified explicitly that the inclusion of the spin-5/2 
resonances mentioned above do not help eliminate the problem of the 
double-bump structure in the predicted $K^+\Xi^-$ invariant mass distribution 
mentioned in the Introduction.

The interaction Lagrangian and propagators of higher spin resonances
that are needed to construct the respective production amplitudes
[cf. diagrams (a)--(g) in Figs.~\ref{fig:diagram}
with $Y=Y'$ in (d) and $\Xi=\Xi'$ in (e)] are given in the Appendix.
In this work, we adopt the pseudovector coupling for the interactions of spin-$1/2$
resonances with the nucleon and with the $\Xi$.

\section{Results}

We now present the numerical results for the observables in the reaction of
$\gamma p \to K^+ K^+ \Xi^-$. 
To be specific, as mentioned above,
we adopt the pseudovector coupling for the meson couplings 
of spin-$1/2$ particles.
As in Ref.~\cite{NOH06}, we include the $\Lambda(1800)\frac12^-$ and 
$\Lambda(1890)\frac32^+$ on top of the $\Lambda(1116)$, $\Lambda(1405)$, 
$\Lambda(1520)$, $\Sigma(1190)$, and $\Sigma(1385)$ hyperons.
In addition, we include the hyperon resonances with masses around $2$~GeV. 
Table~\ref{tbl:hyperons} shows that there are several $\Lambda$ and 
$\Sigma$ resonances in this mass region, but most of them are expected to have 
small contributions according to our discussion in the previous section, as 
their spin-parity quantum numbers are $\frac32^-$, $\frac52^+$, and $\frac72^-$.
The only exception is the $\Sigma(2030)$ with the spin-parity of $\frac72^+$. 
Therefore, to reduce the number of uncontrollable free parameters, 
we include only this resonance in the model of Ref.~\cite{NOH06}.

%%%%%%%%%%%%%%%%%%%%%%%%%%%%%%%%%%%%%%%%%%%%%%
%    Table (II)
\begin{table*}[t]
%\begin{table}[ht]
\begin{center}
\caption{The parameter values employed in the present calculation including the 
$\Sigma(2030)\frac72^+$ resonance. 
The other parameter values are kept the same as given in Ref.~\cite{NOH06}.}
\begin{tabular}{l@{\qquad}r}
\hline\hline
Free parameters                                & Values         \\
\hline\hline
Signs of coupling constants:                        &           \\
$g_{B\Lambda K}^{}$ for $\Lambda(1405)$        & $0.91$     \\
$g_{\Xi\Lambda K}^{}$ for $\Lambda(1405)$      & $0.91$     \\
$g_{\Lambda\Lambda'\gamma}^{}$ for $\Lambda(1116)\leftrightarrow\Lambda(1520)$ transition  
& $1.26$ \\
$g_{\Sigma\Lambda'\gamma}^{}$ for $\Lambda(1193)\leftrightarrow\Lambda(1520)$ transition  
& $2.22$ \\
\hline
Baryonic form factor:                                &          \\
All exponents $n$                                    & $\infty$ \\
$\Lambda_B$ for $p$                                  & 1250MeV  \\
$\Lambda_B$ for $\Xi$                                & 1250MeV  \\
$\Lambda_B$ for spin-1/2 hyperons                    & 1250MeV  \\
$\Lambda_B$ for spin-3/2 hyperons                    & 1230MeV   \\
$\Lambda_B$ for spin-7/2 hyperons                    & 800MeV   \\
\hline
Product of coupling constants:                       &          \\
$g_{N\Lambda K}g_{\Xi\Lambda K}$ for $\Lambda(1800)$ & $1.5$  \\
$g_{N\Lambda K}g_{\Xi\Lambda K}$ for $\Lambda(1890)$ & $2.0$  \\
$g_{N\Sigma K}g_{\Xi\Sigma K}$ for $\Sigma(2030)$    & $-4.2$ \\
\hline\hline
\end{tabular}
\end{center}
\label{tbl:cc}
\end{table*}
%
%%%%%%%%%%%%%%%%%%%%%%%%%%%%%%%%%%%%%%%%%%%%%%%%%%%%%%%%%%%%%%%%%%%

The propagator and effective Lagrangian of spin-$7/2$ resonance are given explicitly 
in the Appendix.
It also contains those of spin-$5/2$ resonance for completeness.
Since we are considering the reaction of $\gamma p \to K^+ K^+ \Xi^-$ in the present work, the
intermediate hyperon resonance has neutral charge. Furthermore, we have no information on the
magnetic moment of $\Sigma(2030)$ resonances. Therefore, we do not include the diagram of 
Fig.~\ref{fig:diagram}(d) corresponding to this resonance in the present calculation.
The values of the free parameters used in the present calculation are given 
in Table~\ref{tbl:cc}.
The other parameter values of the model are given in Ref.~\cite{NOH06}.

\begin{figure*}[t!]
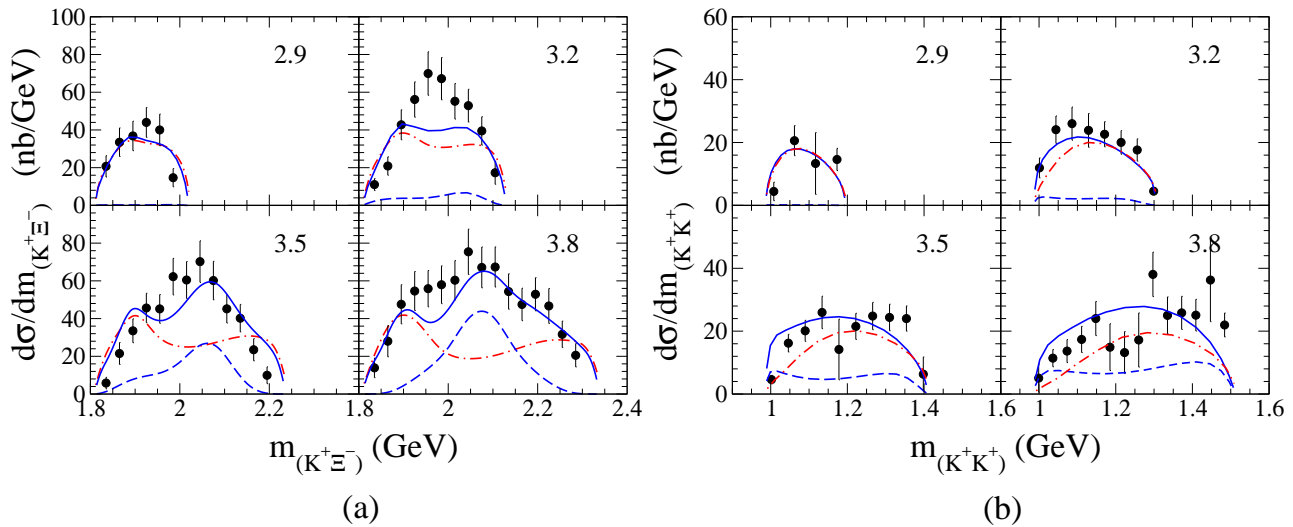
\centering
\vskip -0.4cm
\includegraphics[width=0.47\textwidth,angle=0,clip]{fig2a.eps}
\includegraphics[width=0.47\textwidth,angle=0,clip]{fig2b.eps}
\caption{\label{fig:mkc}
Invariant mass distribution (a) of the $K^+ \Xi^-$ pair and (b) of the $K^+K^+$ pair
in the reaction of $\gamma p \to K^+ K^+ \Xi^-$.
The number in the right upper corner of each graph indicates the incident
photon energy in GeV. The dot-dashed lines are the results of Ref.~\cite{NOH06} which
includes only spin-$1/2$ and -$3/2$ hyperon resonances. The solid lines are the 
results of the present model, while the dashed lines show
the contributions from the $\Sigma(2030)7/2^+$.  
Experimental data are from Ref.~\cite{CLAS07b}.}
\end{figure*}

Our results for the invariant mass distributions of the $K^+ \Xi^-$ pair
and of the $K^+K^+$ pair are given in Fig.~\ref{fig:mkc}.
In Fig.~\ref{fig:mkc}, the dot-dashed lines are the results of Ref.~\cite{NOH06}, where
only the $\Lambda(1800)1/2^-$ and $\Lambda(1890)3/2^+$ are included on top of the
low-lying resonances.
In contrast to the experimental data from the CLAS collaboration~\cite{CLAS07b},
this model yields a double-bump structure in the $K^+\Xi^-$ invariant mass distribution
as illustrated in Fig.~\ref{fig:mkc}(a),
which evidently shows that the contribution from a resonance with a mass around 2~GeV is
missing.
As can be seen by the dashed lines in Fig.~\ref{fig:mkc}(a), 
this can be attributed to the role of the $\Sigma(2030)$.
When combined with the other terms, the double-bump structure present in the model of 
Ref.~\cite{NOH06} now disappears as shown by the solid lines in Fig.~\ref{fig:mkc}(a).
This leads us to conclude that the $\Sigma(2030)$ of $J^P = \frac72^+$ 
has a key role to bring the model predictions into a fair agreement with the CLAS data.
As expected, the contribution from the $\Sigma(2030)7/2^+$ resonance can hardly be seen at low
energies with $E_\gamma \le 3.2$~GeV. Therefore, the results of the present model are
very similar to those of the model of Ref.~\cite{NOH06} at these energies.
However, at higher energies, the range of the invariant $K^+\Xi^-$ mass fully covers 
the mass region of the $\Sigma(2030)$ and the results show the prominent contribution from
this resonance.

%%%%%%%%%%%%%%%%%%%%%%%%%%%%%%%%%%%%%%%%%%%%%%
%    Figure 3
%
%%%%%%%%%%%%%%%%%%%%%%%%%%%%%%%%%%%%%%%%%%%%%%%%%

Figure~\ref{fig:mkc}(b) shows the results for the invariant mass distribution of the
$K^+K^+$ pair in the reaction of $\gamma p \to K^+ K^+ \Xi^-$.
Here again, the contribution from the $\Sigma(2030)$ resonance is large for higher
photon energies. However, as in the model of Ref.~\cite{NOH06}, the obtained $K^+K^+$
invariant mass distribution does not show any structure as expected from the absence of 
exotic $S=+2$ mesons.

%%%%%%%%%%%%%%%%%%%%%%%%%%%%%%%%%%%%%%%%%%%%%%
%    Figure 4
\begin{figure*}[t!]
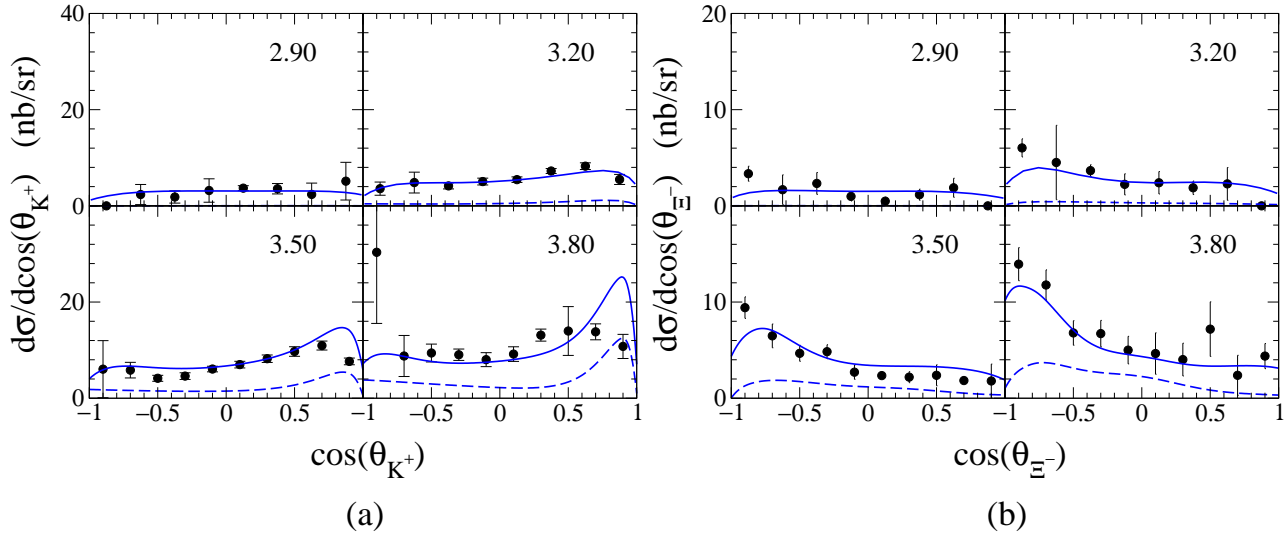
\centering
\vskip -0.4cm
\includegraphics[width=0.47\textwidth,angle=0,clip]{fig3a.eps}
\includegraphics[width=0.47\textwidth,angle=0,clip]{fig3b.eps}
\caption{\label{fig:diff-cs}
Differential cross section (a) $d\sigma/d\cos(\theta_{K^+})$ and 
(b) $d\sigma/d\cos(\theta_{\Xi^-})$ for $\gamma p \to K^+ K^+ \Xi^-$.
The number in the right upper corner of each graph indicates the incident
photon energy in GeV. The solid lines are the full
results of the present model, while the dashed lines show
the contributions from the $\Sigma(2030)7/2^+$. 
Experimental data are from Ref.~\cite{CLAS07b}.}
\end{figure*}
%
%%%%%%%%%%%%%%%%%%%%%%%%%%%%%%%%%%%%%%%%%%%%%%%%%

In Fig.~\ref{fig:diff-cs}, we present the results for the differential cross sections
$d\sigma/d\cos(\theta_{K^+})$ and $d\sigma/d\cos(\theta_{\Xi^-})$ for $\gamma p \to K^+ K^+ \Xi^-$.
This again shows that the role of the $\Sigma(2030)$ becomes prominent as the incident
photon energy increases and is crucial to reproduce the measured data for differential cross sections.
In particular, $d\sigma/d\cos(\theta_{K^+})$ is forward-peaked and $d\sigma/d\cos(\theta_{\Xi^-})$
is backward-peaked, which supports the analyses of Ref.~\cite{NOH06} and is verified by 
the experimental data. The contribution from the $\Sigma(2030)$
also respects this behavior.

\section{Summary and Discussion}

In the present paper, we have investigated the role of hyperon resonances with a mass
around 2~GeV to understand the mechanism of $\Xi^-$ photoproduction.
Our model improves the previous one of Ref.~\cite{NOH06} by including the $\Sigma(2030)$
as a high mass resonance.
Since most of the resonances in this mass region have a spin higher than 5/2, 
we first developed a formalism for
describing the interactions and decays of spin-5/2 and -7/2 resonances based on the
Rarita-Schwinger formalism.
Then, by analyzing the production amplitude near the mass shell, we found that the 
$J^P = \frac52^-$ and $\frac72^+$ resonances would have nontrivial role in 
the production mechanism.
In the mass region of 2~GeV, the only candidate among the hyperon resonances listed 
in the PDG is the $\Sigma(2030)$, and we have included this resonance 
in the present extended version of the model of Ref.~\cite{NOH06}.
In Ref.~\cite{NOH06}, a double-bump structure has been predicted
in the $K^+\Xi^-$ invariant mass distribution in $\gamma p \to K^+ K^+ \Xi^-$,
which is in marked discrepancy with the experimental data \cite{CLAS07b}. 
It has been suggested \cite{NOH06} that this discrepancy might be due
to the missing of a resonance with a mass around 2~GeV.
In this paper, we have explicitly shown that the $\Sigma(2030)$ has a crucial role
in eliminating this double-bump structure.

Since there are many hyperon resonances which can contribute to $\Xi^-$ photoproduction,
the fitted coupling constants should be regarded as a very rough estimate and the analyses
based on more sophisticated models are required to obtain more reliable values for the
resonance parameters as well as more precise experimental data for various physical quantities
of this reaction. 
However, our results show that the high-spin resonances, which have
a mass of around 2~GeV, have a crucial role in the mechanism of $\Xi^-$ photoproduction.
In addition, the formalism for high-spin resonances discussed in the present work can be
applied to analyze other reaction processes.

\acknowledgments

We are grateful to L.~Guo and T.-S.~H. Lee for fruitful discussions.
We also thank V.~Pascalutsa and O.~Scholten for discussions on high-spin
fields.
Y.O. is grateful to the Excited Baryon Analysis Center of the Thomas Jefferson
National Accelerator Facility, where part of this work was done.
We also acknowledge the University of Georgia Research Computing Center for
providing the necessary computing resources.
This work was supported by the FFE Grant \mbox{No.} 41788390
(COSY-58) and by Basic Science Research Program through the National
Research Foundation of Korea (NRF) funded by the Ministry of Education,
Science and Technology (Grant \mbox{No.} 2010-0009381).

\appendix*

\section{\boldmath Formalism of baryon of spin-$5/2$ and -$7/2$}

We refer the details for the interaction Lagrangian and propagators of
spin-$1/2$ and -$3/2$ hyperons to Ref.~\cite{NOH06}.
Here we present our formalism for spin-$5/2$ and -$7/2$ hyperons.

\subsection{Propagators}

Following Refs.~\cite{BF57,Rush66,Chang67a}, we adopt the
Rarita-Schwinger method to describe spin-$5/2$ and -$7/2$ baryon fields.
Then the spin-$5/2$ field is described by a rank-$2$ tensor
$R_{\mu\nu}^{}$ and the spin-$7/2$ field by a rank-$3$ tensor
$R_{\mu\nu\lambda}^{}$.
The propagators of the spin-$5/2$ and -$7/2$ fermion fields are written as
\begin{eqnarray}
S_{5/2}(p) &=& \frac{i}{\slashed{p} - m_R^{} + i \Gamma/2}
\Delta_{\alpha_1^{} \alpha_2^{}}^{\beta_1^{} \beta_2^{}},
\nonumber \\
S_{7/2}(p) &=& \frac{i}{\slashed{p} - m_R^{} + i \Gamma/2}
\Delta_{\alpha_1^{} \alpha_2^{} \alpha_3^{}}^{\beta_1^{} \beta_2^{}
\beta_3^{}},
\end{eqnarray}
where $m_R^{}$ and $\Gamma$ are the mass and the decay width of the
resonance, respectively.
The spin projection operators are obtained by~\cite{Fro58}
\begin{equation}
\sum_{\rm spins} R_{\alpha_1^{} \alpha_2^{} \dots} \bar{R}^{\beta_1^{}
\beta_2^{} \dots} = \Lambda_\pm \Delta_{\alpha_1^{} \alpha_2^{}
\dots}^{\beta_1^{} \beta_2^{} \dots} ,  
\end{equation}
where $\Lambda_\pm$ are the usual energy projection operators of the
spin-$1/2$ theory.
Explicitly, they are obtained as~\cite{BF57,Rush66,Chang67a}
\begin{widetext}
\begin{eqnarray}
\Delta_{\alpha_1^{}\alpha_2^{}}^{\beta_1^{}\beta_2^{}}
({\textstyle\frac52}) &=&
\frac12 \left( \theta_{\alpha_1^{}}^{\beta_1^{}}
\theta_{\alpha_2^{}}^{\beta_2^{}}
+ \theta_{\alpha_1^{}}^{\beta_2^{}}
\theta_{\alpha_2^{}}^{\beta_1^{}} \right)
-\frac15 \theta_{\alpha_1^{}\alpha_2^{}} \theta^{\beta_1^{}\beta_2^{}}
%\nonumber \\ && \mbox{}
- \frac{1}{10} \left(
\bar{\gamma}_{\alpha_1^{}} \bar{\gamma}^{\beta_1^{}}
\theta_{\alpha_2^{}}^{\beta_2^{}}
+ \bar{\gamma}_{\alpha_1^{}} \bar{\gamma}^{\beta_2^{}}
\theta_{\alpha_2^{}}^{\beta_1^{}}
+ \bar{\gamma}_{\alpha_2^{}} \bar{\gamma}^{\beta_1^{}}
\theta_{\alpha_1^{}}^{\beta_2^{}}
+ \bar{\gamma}_{\alpha_2^{}} \bar{\gamma}^{\beta_2^{}}
\theta_{\alpha_1^{}}^{\beta_1^{}} \right).
\nonumber \\
\Delta_{\alpha_1^{}\alpha_2^{}\alpha_3^{}}
^{\beta_1^{}\beta_2^{}\beta_3^{}} ({\textstyle\frac72}) &=& 
\frac{1}{36} \sum_{P(\alpha),P(\beta)} \left\{ 
\theta_{\alpha_1^{}}^{\beta_1^{}} \theta_{\alpha_2^{}}^{\beta_2^{}}
\theta_{\alpha_3^{}}^{\beta_3^{}}
-\frac37
\theta_{\alpha_1^{}}^{\beta_1^{}} \theta_{\alpha_2^{}\alpha_3^{}}
\theta^{\beta_2^{}\beta_3^{}}
-\frac37
\bar{\gamma}_{\alpha_1^{}} \bar{\gamma}^{\beta_1^{}}
\theta_{\alpha_2^{}}^{\beta_2^{}} \theta_{\alpha_3^{}}^{\beta_3^{}}
%\right. \nonumber \\  && \left. \mbox{} \qquad \qquad \qquad
+ \frac{3}{35}
\bar{\gamma}_{\alpha_1^{}} \bar{\gamma}^{\beta_1^{}}
\theta_{\alpha_2^{}\alpha_3^{}} \theta^{\beta_2^{}\beta_3^{}} \right\},
\label{eq:prop}
\end{eqnarray}
\end{widetext}
where
\begin{eqnarray}
\theta_\mu^\nu = g_\mu^\nu - \frac{p_\mu p^\nu}{M^2}, \qquad
\bar\gamma_\mu = \gamma_\mu - \frac{p_\mu \slashed{p}}{M^2}
\end{eqnarray}
with $M$ being the resonance mass.
In Eq.~(\ref{eq:prop}), the propagator for spin-$7/2$ field contains
summation over all possible permutations of
$\{\alpha_1^{}, \alpha_2^{}, \alpha_3^{} \}$ 
and of $\{\beta_1^{}, \beta_2^{}, \beta_3^{} \}$.

\subsection{Effective Lagrangian}

By considering parity and angular momentum conservation, one can easily
confirm that there is only one form factor in the interaction of
a hyperon of arbitrary spin with the nucleon-kaon channel or with the
$\Xi$-kaon channel. 
By introducing
\begin{equation}
\Gamma^{(\pm)} = \left( \begin{array}{c} \gamma_5^{} \\ 1 \end{array}
\right),
\end{equation}
and dropping the isospin dependence,
the interaction Lagrangian of spin-$5/2$ hyperon field $Y_{\mu\nu}$
reads
\begin{eqnarray}
\mathcal{L}_{BYK}^{5/2^\pm} &=&
i \frac{g_{BYK}^{}}{m_K^2} \bar{B} \Gamma^{(\pm)}
Y^{\mu\nu} \partial_\mu \partial_\nu \bar{K} + \mbox{ H.c.},
\nonumber \\
\mathcal{L}_{BYK\gamma}^{5/2^\pm} &=& - \frac{g_{BYK}^{}}{m_K^2}
\hat{e}_K^{}
\bar{B} \Gamma^{(\pm)} Y^{\mu\nu} \left( A_\mu \partial_\nu +
\partial_\nu A_\nu \right) \bar{K} 
\nonumber \\
&& \mbox{} + \mbox{ H.c.},
\end{eqnarray}
where the latter is obtained from the former by minimal substitution.
Here, $\hat{e}_K$ is the charge of the $K$ meson.

The corresponding Lagrangian for a spin-$7/2$ resonance are obtained as
\begin{eqnarray}
\mathcal{L}_{BYK}^{7/2^\pm} &=& - \frac{g_{BKY}^{}}{m_K^3} \bar{B}
\Gamma^{(\mp)} Y^{\mu\nu\alpha} \partial_\mu \partial_\nu \partial_\alpha
K + \mbox{ H.c.},
\nonumber \\
\mathcal{L}_{BYK\gamma}^{7/2^\pm} &=& i \frac{g_{BYK}^{}}{m_K^3} \bar{B}
\Gamma^{(\pm)} Y^{\mu\nu} \hat{e}_K A_{\mu\nu} \bar{K} + \mbox{ H.c.}.
\end{eqnarray}

\end{document}